\documentclass[prl,nofootinbib,twocolumn,superscriptaddress,showpacs,showkeys]{revtex4-1}

\usepackage{graphicx,epsfig}
\usepackage{amsfonts,amsmath,amssymb,amsthm,amscd}
\usepackage{color}
\usepackage{dcolumn}
\usepackage{bm}
\usepackage{array}
\usepackage{float}
\usepackage[]{units}
\usepackage[colorlinks=true,linkcolor=blue,citecolor=blue]{hyperref}

\renewcommand\vec{\mathbf}

\begin{document}

\title{Multiple-colliding laser pulses as a basis for studying high-field high-energy physics}

\author{J. Magnusson}
\email{joel.magnusson@chalmers.se}
\affiliation{Department of Physics, Chalmers University of Technology, SE-41296 Gothenburg, Sweden}

\author{A. Gonoskov}
\affiliation{Department of Physics, University of Gothenburg, SE-41296 Gothenburg, Sweden}
\affiliation{Institute of Applied Physics, Russian Academy of Sciences, Nizhny Novgorod 603950, Russia}

\author{M. Marklund}
\affiliation{Department of Physics, University of Gothenburg, SE-41296 Gothenburg, Sweden}

\author{T. Zh. Esirkepov}
\affiliation{Kansai Photon Science Institute, National Institutes for Quantum and Radiological
Science and Technology (QST), 8-1-7 Umemidai, Kizugawa, Kyoto 619-0215, Japan}

\author{J. K. Koga}
\affiliation{Kansai Photon Science Institute, National Institutes for Quantum and Radiological
Science and Technology (QST), 8-1-7 Umemidai, Kizugawa, Kyoto 619-0215, Japan}

\author{K. Kondo}
\affiliation{Kansai Photon Science Institute, National Institutes for Quantum and Radiological
Science and Technology (QST), 8-1-7 Umemidai, Kizugawa, Kyoto 619-0215, Japan}

\author{M. Kando}
\affiliation{Kansai Photon Science Institute, National Institutes for Quantum and Radiological
Science and Technology (QST), 8-1-7 Umemidai, Kizugawa, Kyoto 619-0215, Japan}

\author{S. V. Bulanov}
\affiliation{Kansai Photon Science Institute, National Institutes for Quantum and Radiological
Science and Technology (QST), 8-1-7 Umemidai, Kizugawa, Kyoto 619-0215, Japan}
\affiliation{Institute of Physics ASCR, v.v.i. (FZU), ELI-Beamlines Project, 182 21 Prague, Czech Republic}
\affiliation{Prokhorov General Physics Institute RAS, Vavilov Str. 38 , Moscow 119991, Russia}

\author{G. Korn}
\affiliation{Institute of Physics ASCR, v.v.i. (FZU), ELI-Beamlines Project, 182 21 Prague, Czech Republic}

\author{C. G. R. Geddes}
\affiliation{Lawrence Berkeley National Laboratory, Berkeley, California 94720, USA}

\author{C. B. Schroeder}
\affiliation{Lawrence Berkeley National Laboratory, Berkeley, California 94720, USA}

\author{E. Esarey}
\affiliation{Lawrence Berkeley National Laboratory, Berkeley, California 94720, USA}

\author{S. S. Bulanov}
\affiliation{Lawrence Berkeley National Laboratory, Berkeley, California 94720, USA}

\begin{abstract}
Apart from maximizing the strength of optical electromagnetic fields achievable at high-intensity laser facilities, the collision of several phase-matched laser pulses has been theoretically identified as a trigger of and way to study various phenomena. These range from the basic processes of strong-field quantum electrodynamics to the extraordinary dynamics of the generated electron-positron plasmas. This has paved the way for several experimental proposals aimed at both fundamental studies of matter at extreme conditions and the creation of particle and radiation sources. Because of the unprecedented capabilities of such sources they have the potential to open up new opportunities for experimental studies in nuclear and quark-gluon physics. We here perform a systematic analysis of different regimes and opportunities achievable with the concept of multiple-colliding laser pulses (MCLP), for both current and upcoming laser facilities. We reveal that several distinct regimes could be within reach of multi-PW laser facilities.\\
\end{abstract}

\maketitle

\section{Introduction}
Studies of the interaction between charged particles with electromagnetic (EM) fields of high intensity has unveiled a large number of phenomena and has received a growing attention over the last decade \cite{mourou.rmp.2006,marklund.rmp.2006,dipiazza.rmp.2012}. This attention is not only due to the diversity of phenomena enabled by strong fields but also because of the emerging experimental capabilities that, apart from collider-type experiments \cite{bula.prl.1996, burke.prl.1997, cole.prx.2018, poder.prx.2018}, will provide opportunities for performing experiments in many other configurations. The next generation of laser facilities, such as ELI Beamlines, ELI NP, CoReLS, Apollon, XCELS, and Vulcan-10\,PW \cite{ELI-beamlines, ELI-NP, CORELS, APOLLON, XCELS, Vulcan}, as well as future lepton colliders \cite{DOE.Roadmap} are expected to operate in a regime where these phenomena either dominate or significantly influence the interaction. 

According to quantum electrodynamics (QED), the action of strong EM fields can lead to the photon emission of an electron (Compton process \cite{brown.pr.1964,nikishov.jetp.1964,goldman.pl.1964}), the photon decay into electron-positron pairs (Breit-Wheeler process \cite{nikishov.jetp.1964,breit.pr.1934,reiss.jmp.1962}), and the electron-positron pair production from vacuum (Schwinger process \cite{sauter.zphys.1931, heisenberg.zphys.1936, schwinger.pr.1951}). In forthcoming experiments the field intensity can be so high that the two former processes appear in rapid succession, leading to a cascaded production of electrons, positrons and high-energy photons \cite{bell.prl.2008, fedotov.prl.2010, bulanov.prl.2010b}. Apart from the shower-type cascade, which implies a repeated divison of the energy of the initial particle or photon, high-intensity lasers also open up the possibility for running avalanche-type cascades, in which the electromagnetic fields both accelerate particles and cause the QED processes \cite{fedotov.prl.2010, bulanov.pra.2013}. Such cascade can be initiated by the seed particles of energetic particle beams, a background plasma or even by the Schwinger process. These cascades are fascinating phenomena of rapid transformation of laser and charged particle beam energy into high-energy photons and electron-positron pairs. 

In order to trigger these processes one needs to reach certain strength of the electromagnetic field. This raises the problem of finding the optimal strategy of focusing laser radiation. One such strategy, first found to be advantageous for enhancing the electron-positron pair production, is the concept of multiple colliding laser pulses (MCLP) \cite{bulanov.prl.2010a}. In this concept, a laser beam is split into $N$ equal sub-beams that are subsequently combined constructively at focus. The energy of the un-split laser beam $\mathcal{E}_1$ is related to its peak electric field $E_1$ and peak intensity $I_1$ as $\mathcal{E}_1\sim E_1^2\sim I_1$. Each sub-beam receives an energy of $\mathcal{E}_1/N$, and, with identical focusing, has a peak electric field of $E_1/\sqrt{N}$ and intensity of $I_1/N$. A constructive interference of the sub-beams therefore gives an electric field of $E_N=\sqrt{N}E_1$, and an intensity of $I_N=N I_1$, which can be significantly higher than for a single un-split beam. However, for a large number of sub-beams the peak value of the electric field at focus $E_N$ is obviously constrained by the diffraction limit. The problem of maximizing the field strength is limited by the optimal case, which can be viewed as the inversed emission of a dipole antenna. Therefore, the resulting field structure is often also referred to as a dipole wave and represents an accessible form for theoretical analysis \cite{bulanov.prl.2010b, gonoskov.pra.2012, gonoskov.prl.2013}.   

For monochromatic fields, the dipole wave has been shown to provide the highest possible field strength $a_{0} \approx 780 \sqrt{P/\mathrm{PW}}$ for a given power $P$ \cite{bassett.oa.1986} and it has also been shown that this field configuration is beneficial for the generation of QED cascades \cite{gelfer.pra.2015,vranic.ppcf.2016, gonoskov.prx.2017} (the field strength is hereafter written in relativistic units that will be rigorously defined later). Apart from enhancing pair production due to the Breit-Wheeler process the dipole wave also has the interesting property of trapping charged particles near the maxima of its electric field, for sufficiently high field intensities \cite{gonoskov.prl.2014}. This is referred to as \emph{anomalous} radiative trapping in order to differentiate from the normal radiative trapping that occurs at the magnetic(electric) field maxima(minima) due to gyration of electrons that rapidly exhaust their energy through radiation losses \cite{kirk.ppcf.2016}. It is noteworthy however that anomalous radiative trapping not only occurs in dipole waves, but is a more universal feature that originates from the general tendency of charged particles to align their motion along the radiation-free direction \cite{gonoskov.pop.2018} when they rapidly lose energy and enter into the radiation-dominated regime \cite{bulanov.ppr.2004, bulanov.prl.2010b}. This can even be observed in simple mechanical systems with strong friction forces \cite{esirkepov.pla.2017}. It has also been recently shown that charged particles in the strong EM fields formed by two colliding pulses tend to move along trajectories that are defined by either attractors or limit cycles \cite{fedotov.pra.2014, kirk.ppcf.2016, jirka.pre.2016, gong.pre.2017, samsonov.pra.2018}. This analysis was later expanded to the case of MCLP, for which similar phase space patterns could be observed \cite{bulanov.jpp.2017}. While such behavior is mostly attributed to intense radiation losses that are typically modeled based on the classical description of radiation reaction, in terms of either a Landau-Lifshitz (LL) equation of motion or a ``modified" LL equation \cite{esirkepov.pla.2015,bulanov.jpp.2017}, the mechanisms behind the observed phenomena are tolerant to the quantized nature of emission and similar behaviour has also been observed in computer simulations based on probabilistic quantum treatment of radiation losses \cite{gonoskov.prl.2014, gonoskov.pop.2018, fedotov.pra.2014, jirka.pre.2016}.

A more complete treatment based on simulations that also account for pair production and therefore QED cascades \cite{nerush.prl.2011, duclous.ppcf.2011, sokolov.pop.2011, ridgers.jcp.2014, gonoskov.pre.2015, niel.pre.2018, derouillat.cpc.2018} has helped reveal a fortunate interplay between the cascade and anomalous radiative trapping in the focus of a dipole wave. The trapping mechanism causes a migration of particles to the focus where the strong field makes them gain energy and emit high-energy photons that in turn have a high chance of decaying into a pair of particles. Even if the original particle leaves the system due to insufficient radiation losses, the generated particles of lower energy might be trapped, replenishing the overall number of particles undergoing the trapped motion. This leads to the triggering of self-sustained, locally confined QED cascades for laser powers as low as $P = \unit[7.2]{PW}$ \cite{gonoskov.prx.2017}, which is drastically lower than the $\unit[60]{PW}$ required for a prominent appearance of anomalous radiative trapping in single-particle dynamics \cite{gonoskov.prl.2014}. For $P \gtrsim \unit[8]{PW}$ the exponential growth of the number of particles in such a trapped QED cascade makes it possible to reach extreme states of electron-positron plasma in terms of both high density and high energy of particles and photons. Moreover, the self-action of the tremendous current produced by counter-propagating electrons and positrons in the oscillating dipole field leads to the growth of an instability followed by a collapse into ultrathin radial sheets \cite{efimenko.scirep.2018} and, for $P > \unit[20]{PW}$, into an axial pinch \cite{efimenko.pre.2019}. 

\begin{figure}[t!]
\includegraphics[width=0.5\textwidth]{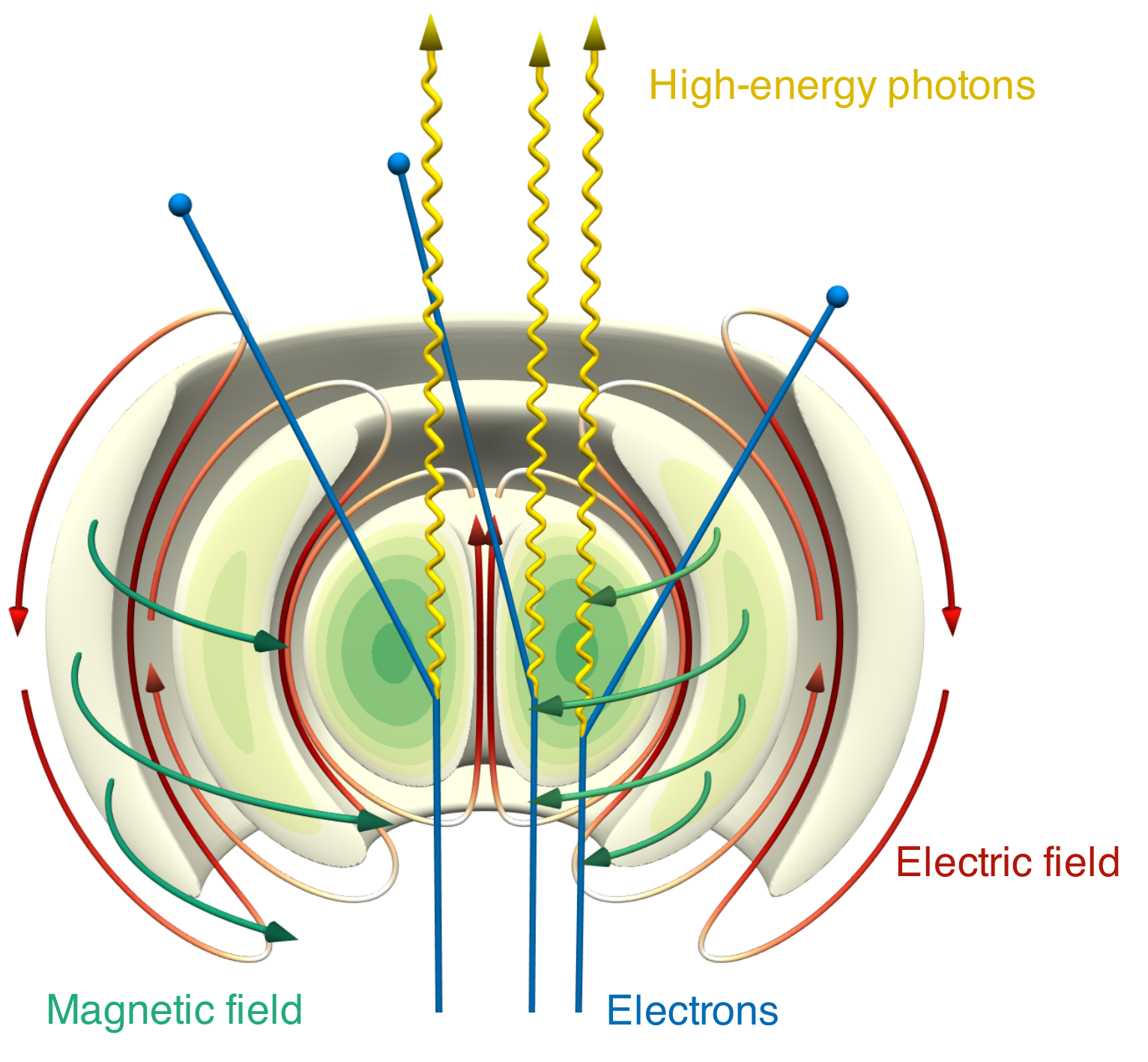}
\caption{Conceptual visualization of the setup, where high energy electrons (blue) are injected along the axis of an intense dipole wave. In this field, the electrons will emit large amounts of high-energy photons (yellow). The polarization of the shown field is that of an electric dipole, with a poloidal electric field (red) and a toroidal magnetic field (green). Figure is reprinted from \cite{magnusson.arxiv.2018}.}
\label{fig:setup}
\end{figure} 

The trapping of charged particles at the maxima of strong EM fields should result in an enhanced emission of high-energy photons. Indeed, it has been shown that in the interaction of a dipole wave with an underdense plasma, a fraction of the plasma electrons gets trapped at the focus and oscillate along the symmetry axis of the wave, producing a bidirectional source of GeV photons \cite{gonoskov.prx.2017}. However, the maximum quiver energy of an electron in this field is $\gamma\sim 2a_0$, which implies a maximum photon energy of $\hbar\omega \sim 0.8 \sqrt{P/\mathrm{PW}}\,\text{GeV}$. This limits the energy of the emitted photons to a few GeV for upcoming $\unit[10]{PW}$ facilities. In order to produce photon energies beyond this limit, in the tens of GeV, the interaction of a highly energetic electron beam with a dipole wave was suggested in Ref.~\cite{magnusson.arxiv.2018}. In this configuration it is interesting to note that, above a certain threshold, increasing the laser power will not lead to a more efficient generation of high-energy photons. It will instead drive an increasingly stronger shower cascade, hampering the yield of high-energy photons. In fact, this was shown to also be the case for more general field structures and, while allowing for a higher efficiency, is therefore not exclusive to the dipole wave. For the purpose of efficiently generating photons above a few GeV it was found that there is an optimal laser power of $P \approx\unit[0.4]{PW}$ for a dipole wave, around which $\sim20\%$ of the initial electrons can be converted into high-energy photons\footnote{We here define \emph{high-energy photons} as those carrying an energy equal to or larger than half the initial electron energy.}. 

Altogether, these results show the flexibility of the MCLP configuration and thereby the possibility of studying several distinct regimes of interaction, depending primarily on values of laser power and electron beam energy. 

In this paper we study the interaction of an electron beam with MCLP, where the beam is directed along the symmetry axis of the MCLP configuration. This allows the electrons to reach the highest EM field strengths without a significant loss of energy or lateral scattering. The electron beam can therefore remain well collimated until it reaches the focus of the MCLP configuration (see Fig.~\ref{fig:setup}), maximizing the chance of trapping electrons in the EM field as they are exposed to the highest intensity. It also opens up a possibility to study the interaction of charged particle with super-critical fields, which is currently an active topic of discussion \cite{blackburn.arxiv.2018,yakimenko.prl.2019,podszus.prd.2019,ilderton.prd.2019}. As mentioned, this setup is also advantageous for high energy photon production for certain values of laser power. We here identify several areas in the (electron beam energy, laser power) - parameter space, favouring the development of shower or avalanche-type cascades, high-energy photon production or electron beam energy-depletion. Thus, the proposed configuration may also provide a solution for effectively stopping high-energy electron beams over micrometer scale distances.

The paper is organized as follows. Section two provides a brief overview of the analytical form of the dipole wave together with analytical estimates for the expected behavior of energetic electrons injected into this field, based on the LL equations of motion. In section three we present the results of numerically modelling the interaction using QED-PIC \cite{gonoskov.pre.2015}. First, we review the results of Ref.~\cite{magnusson.arxiv.2018} on the high-energy photon production occurring at moderate laser powers. Second, we discuss the development of the shower-type cascade at high laser powers and identify different regimes possible within the range of $P < \unit[10]{PW}$ and initial electron energies $< \unit[50]{GeV}$. And, third, we present results on efficient electron beam energy-depletion, based on the analysis of the shower cascade. We conclude in section four and summarize where different regimes of interaction manifest themselves.

\section{Classical motion in a dipole wave field}\label{sec:field}
The EM field of a dipole wave is a three dimensional structure that in general can be described by an expansion in spherical harmonics around its point of focus. It is convenient to express this expansion in terms of transverse magnetic (TM) and transverse electric (TE) modes, which have toroidal magnetic and poloidal electric or poloidal magnetic and toroidal electric fields, respectively. The toroidal magnetic and electric fields can be written in spherical coordinates as \cite{vainshtein.1988}
\begin{equation}
\left(\begin{tabular}{c} $B_\phi$ \\ $E_\phi$ \end{tabular}\right)=
\left(\begin{tabular}{c} $a_\mathrm{TM}\sin t$ \\ $a_\mathrm{TE}\sin\left(t+\varphi_\mathrm{TE}\right)$ \end{tabular}\right) \frac{J_{n+1/2}(R) L^1_n(\cos\theta)}{\sqrt{R}},
\end{equation}           
where $a_\mathrm{TM}$ and $a_\mathrm{TE}$ are the amplitudes of the TM and TE modes respectively, normalized to relativistic units $m_e\omega c/e$, $\varphi_\mathrm{TE}$ is the phase difference between the two modes, $c$ is the speed of light and $m_e$ and $e$ are the electron mass and charge respectively. $J_\nu(x)$ is the Bessel function and $L^l_n(x)$ is the associated Legendre polynomial \cite{abramowitz.1965}. Time $t$ is normalized to $\omega^{-1}$ and distance, $R=\sqrt{x^2+y^2+y^2}$, is normalized to $k=c/\omega$. The poloidal components are readily obtained by making use of the relations between the Fourier components of the electromagnetic fields. The poloidal electric field in the TM mode is thus obtained from $\vec{E} = ik(\nabla\times \vec{B})$ and the poloidal magnetic field in the TE mode from $\vec{B} = -ik(\nabla\times\vec{E})$. For the analysis of electron motion the field is however more conveniently expressed in cylindrical coordinates. For the non-trivial configuration of highest symmetry, with $n=1$, we have
\begin{equation}\label{eq:EM_phi}
\left(\begin{tabular}{c} $B_\phi$ \\ $E_\phi$ \end{tabular}\right)=
\left(\begin{tabular}{c} $a_\mathrm{TM}\sin t$ \\ $a_\mathrm{TE}\sin t$ \end{tabular}\right)\sqrt{\frac{2}{\pi}}\rho\left[\frac{\sin R - R\cos R}{R^3}\right],
\end{equation}
where $\rho=\sqrt{x^2+z^2}$. The poloidal components of the electric and magnetic fields are in turn given by
\begin{equation}\label{eq:EM_rho}
\left(\begin{tabular}{c} $E_\rho$ \\ $B_\rho$ \end{tabular}\right)=
\left(\begin{tabular}{c} $a_\mathrm{TM}\cos t$ \\ $a_\mathrm{TE}\cos t$ \end{tabular}\right)\sqrt{\frac{2}{\pi}}z \rho\left[\frac{3R\cos R + (R^2-3)\sin R}{R^5}\right]
\end{equation}
and
\begin{equation}\label{eq:EM_z}
\begin{aligned}
\left(\begin{tabular}{c} $E_z$ \\ $B_z$ \end{tabular}\right) =
& \left(\begin{tabular}{c} $a_\mathrm{TM}\cos t$ \\ $a_\mathrm{TE}\cos t$ \end{tabular}\right)\sqrt{\frac{2}{\pi}} \\
\times & \left[\frac{(2z^2-\rho^2)R\cos R - (2z^2-\rho^2+\rho^2R^2)\sin R}{R^5}\right].
\end{aligned}
\end{equation}
The maximum of the TM(TE) mode is reached at the centre of the dipole wave, with a value of $\sqrt{8/9\pi}a_\mathrm{TM}$ ($\sqrt{8/9\pi}a_\mathrm{TE}$), and is directed along the dipole axis.

\begin{figure*}[t!]
\includegraphics[width=1\textwidth]{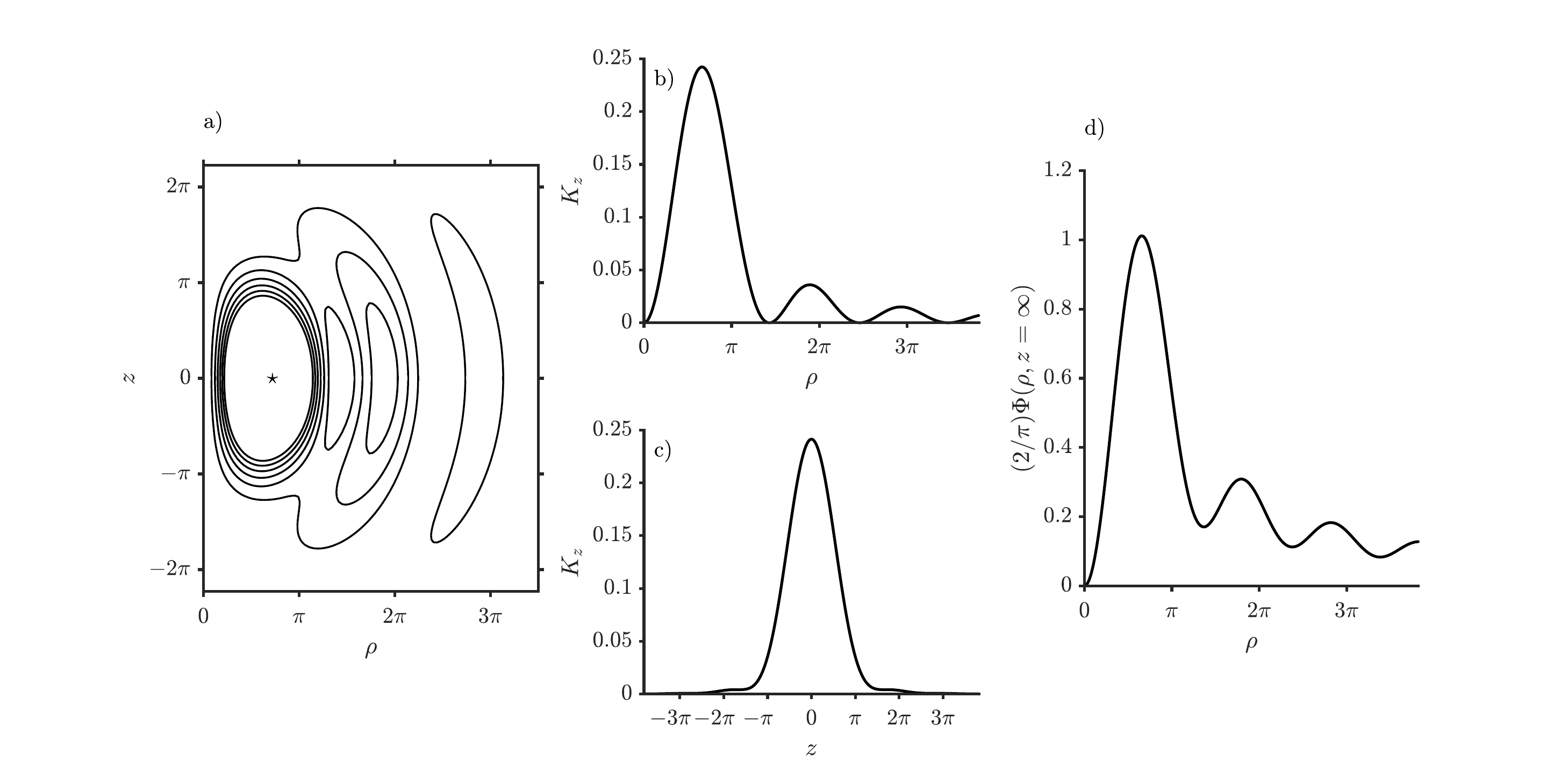}
\caption{(a) Isocontours of the function $K_z$ in the $(\rho,z)$ plane. (b) The dependence of the function $K_z$ on the radial coordinate $\rho$ at $z=0$. (c) The dependence of the function $K_z$ on the longitudinal coordinate $z$ at $\rho=\rho^\star \approx 2$. (d) The dependence of the function $\Phi$ on the radial coordinate $\rho$ at $z=\infty$.}\label{fig:Kz}
\end{figure*}

\subsection{Equations of electron motion}
The motion of an electron in an EM field is described by the following equations
\begin{equation}
\dot{\vec{p}}=\vec{E}+\vec{p}\times\vec{B}/\gamma+\vec{g}_\mathrm{rad},
\quad
\dot{\vec{x}}=\vec{p}/\gamma,
\end{equation}   
where $\vec{x}$ and $\vec{p}$ are the electron position and momentum, respectively, and $\gamma=(1+\vec{p}^2)^{1/2}$ is the Lorentz factor. The momentum $\vec{p}$ is normalized to $m_e c$ and the fields are normalized to $m_e \omega c/e$. The radiation reaction force $\vec{g}_\mathrm{rad}$ can for ultra-relativistic motion be written in the Landau-Lifshitz form \cite{landau.vol2.1971}
\begin{equation}\label{g_rad}
\vec{g}_\mathrm{rad}=-\frac{2\alpha a_S^2\chi_e^2}{3\gamma}\vec{p}.
\end{equation}
Here $\alpha=e^2/\hbar c\approx 1/137$ is the fine structure constant, $a_S=m_e c^2/\hbar \omega$ is the normalized QED critical field, $\hbar$ is the reduced Planck constant, and the nonlinear quantum parameter $\chi_e$ is defined as $\chi_e=\sqrt{|F^{\mu\nu}p_\nu|^2}/a_S$ \cite{nikishov.jetp.1964}. The EM field tensor is defined as $F_{\mu\nu}=\partial_\mu A_\nu-\partial_\nu A_\mu$, where $A_\mu$ is the 4-potential and $p^\nu=(\gamma,\vec{p})$ is the 4-momentum of the electron. In three-dimensional notation the nonlinear quantum parameter reads
\begin{equation}\label{chi_e}
\chi_e=\frac{1}{a_S}\sqrt{(\gamma\vec{E}+\vec{p}\times\vec{B})^2-(\vec{p}\cdot\vec{E})^2}.
\end{equation}  
It follows from Eqs. (\ref{g_rad}-\ref{chi_e}) that the expression for the radiation reaction force can be rewritten as
\begin{equation}
\vec{g}_\mathrm{rad}=-\epsilon_\mathrm{rad}\vec{p}\gamma\left[\left(\vec{E}+\frac{1}{\gamma}\vec{p}\times\vec{B}\right)^2-\left(\frac{1}{\gamma}\vec{p}\cdot\vec{E}\right)^2\right],
\end{equation}
where we for later convenience have introduced the dimensionless parameter $\epsilon_\mathrm{rad}=2r_e/3\lambdabar$ with $r_e = e^2/m_e c^2 \approx \unit[2.82\times 10^{-13}]{cm}$ being the classical electron radius, $\lambdabar$ the reduced laser wavelength, and $\epsilon_\mathrm{rad}=1.2\times 10^{-8}$ for $2\pi\lambdabar=\unit[1]{\mu m}$. The parameter $\epsilon_\mathrm{rad}$ characterizes how the radiation reaction affects the dynamics of a radiating electron.

\begin{figure*}[t!]
	\includegraphics[width=1\textwidth]{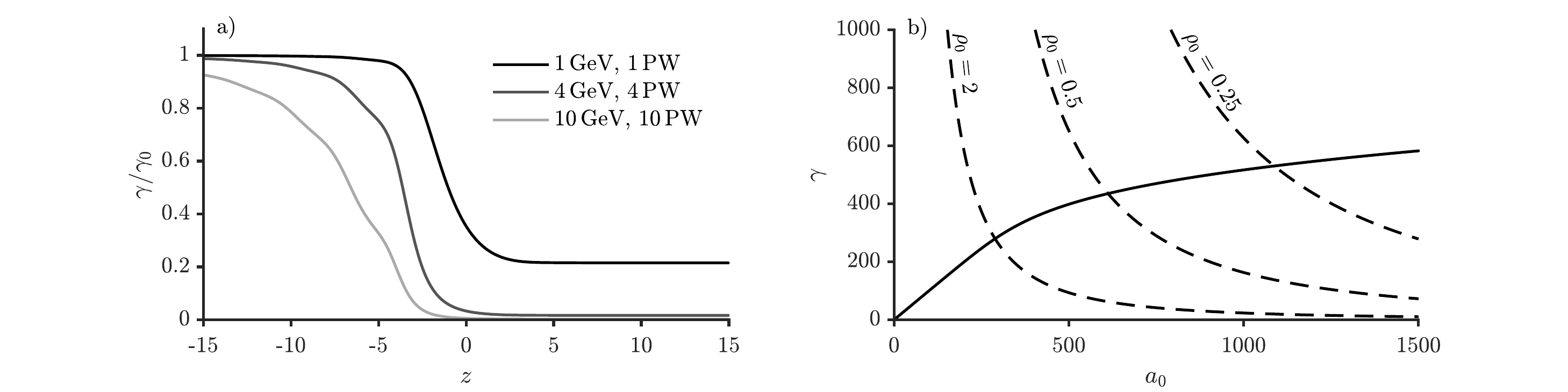}
	\caption{(a) The evolution of the electron energy in the EM dipole field according to equation~\eqref{eq:p_z} for $\rho_0=0.5$ and three sets of initial electron energy and total power. (b) The dependence of the electron energy on peak amplitude of the EM field according to equation~\eqref{p_RR_tr} (solid) and equation~\eqref{p_RR_long} (dashed) for three different values of $\rho_0$.}\label{fig:classical}
\end{figure*}

In order to obtain analytical estimates for the electron energy under the influence of radiation reaction in a strong EM field, we turn to the simple case of an ultrarelativistic electron rotating in the anti-node of a circularly polarized electromagnetic wave. In this scenario, the power emitted is proportional to the forth power of the electron's energy $m_e c^2\omega\epsilon_\mathrm{rad}\gamma^2(\gamma^2-1)$. Simultaneously, the electron can acquire energy from the electromagnetic field at a rate of $m_e c^2\omega a_0$. The steady-state condition of balance between the acquired and lost energy therefore yields $a_0^3\approx \epsilon_\mathrm{rad}^{-1}$ with the result that radiation friction effects will become dominant in the limit of large EM field amplitudes, when $a_0>a_\mathrm{rad}=\epsilon_\mathrm{rad}^{-1/3}$. In this regime, and because of the energy loss, the electron energy is determined by (see Ref. \cite{bulanov.pre.2011}):
\begin{equation}\label{p_RR_tr}
a_0^2=(\gamma^2-1)(1+\epsilon_{rad}^2\gamma^6),~~~\tan\varphi=\left(\epsilon_{rad}\gamma^3\right)^{-1}
\end{equation}
where $\varphi$ is the angle between the directions of the electron momentum and the electric field. For $a_0\gg a_\mathrm{rad}=\epsilon_\mathrm{rad}^{-1/3}$ the energy of an electron therefore scales as 
\begin{equation}\label{eq:energy_estimate}
\varepsilon \sim m_e c^2(a_0/\epsilon_\mathrm{rad})^{1/4}
\end{equation}
which may also serve as a rough estimate for the electron energy in more general EM fields.

\subsection{Electron energy loss due to radiation reaction}\label{sec:classical_energy_loss}

An ultra-relativistic electron crossing a region of strong electromagnetic fields will lose energy due to radiation friction, the rate of which is easily obtained from the equations of motion to be
\begin{equation}
\dot{\gamma}=\frac{1}{\gamma}(\vec{p}\cdot\vec{E}+\vec{p}\cdot\vec{g}_\mathrm{rad}).
\end{equation}
In order to emphasize the effects of radiation reaction we consider the scenario of a strong field $a\gg 1$ such that the longitudinal component of the electron momentum dominates over the transverse component acquired in the EM field, meaning $p \gg a$. Writing the electron momentum as $\vec{p}=p\vec{e}_p$, the momentum equation will take the form
\begin{equation}
\dot{p}=-\epsilon_\mathrm{rad} p^2\left[\left(\vec{E}+\vec{e}_p\times\vec{B}\right)^2-\left(\vec{e}_p\cdot\vec{E}\right)^2\right].
\end{equation}
For an electron moving parallel to the symmetry axis ($z$-axis) of the MCLP configuration described above, the equation can be further simplified to
\begin{equation}\label{eq:eq_motion}
p^\prime =-\epsilon_\mathrm{rad}p^2 K_z(\rho,z),
\end{equation}
where the primed momentum denotes differentiation with respect to the coordinate $z$ and the function $K_z(\rho,z)=\left(E_\rho-B_\phi\right)^2+\left(E_\phi+B_\rho\right)^2$ governs the distribution of the radiation reaction force acting on the ultra-relativistic electron propagating in the $z$-direction. We now assume that the field is an equal mix of the TE and TM mode $a_\mathrm{TE}=a_\mathrm{TM}=a_m$ with no phase difference $\varphi_\mathrm{TE}=0$. The $K_z$ function then takes the form shown in Figure~\ref{fig:Kz}. In particular, the function is equal to zero on the $z$-axis, has a $\rho^2$ scaling for small radii $\rho\ll 1$ and reaches its maximum of $K_z\approx 0.24 a_m^2$ at $\rho=\rho^\star \approx 2$, $z = 0$. The maximum value of the nonlinear quantum parameter $\chi_e$ that can be achieved by an electron moving parallel to the $z$-axis is $0.49\gamma a_m/a_S$. For an electron with initial momentum $\vec{p}_0$ the solution to Eq.~\eqref{eq:eq_motion} is then given by
\begin{equation}\label{eq:p_z}
p(z)=\frac{p_0}{1+(2/\pi)p_0\epsilon_\mathrm{rad}a_m^2\Phi(\rho_0, z)},
\end{equation}
where $(2/\pi)a_m^2\Phi(\rho_0, z)=\int_{-\infty}^{z}dz^\prime K_z(\rho_0, z^\prime)$ (see Fig.~\ref{fig:classical}(a)). Although equation (\ref{eq:eq_motion}) is at the threshold of its applicability for the parameters chosen to plot the curves in Figure~\ref{fig:classical}(a), the evolution of the electron energy shows a rapid depletion and saturation. We may also note that for $(2/\pi)p_0\epsilon_\mathrm{rad}a_m^2\Phi(\rho_0, z)\gg 1$, the momentum $p(z)$ becomes independent of the initial momentum $p_0$, meaning that the particle in this highly dissipative system ``forgets" about its initial conditions and its final momentum will tend to
\begin{equation}\label{p_RR_long}
p_\infty=\frac{\pi}{2\epsilon_\mathrm{rad}a_m^2\Phi(\rho_0,z=\infty)}.
\end{equation}
For the approximations to be valid during the whole interaction process the condition $p(z)\gg a_0$ must be satisfied, leading to the requirement that
\begin{equation} \label{long_motion}
a_0\ll\left(\frac{4}{9\Phi(\rho_0, z)}\right)^{1/3}\epsilon_\mathrm{rad}^{-1/3}.
\end{equation}
If satisfied, radiation reaction effects associated with the transverse motion should not play any significant role in the dynamics.

Based on the results for the final electron energy in the longitudinal \eqref{p_RR_long} and transverse \eqref{p_RR_tr} motion in strong EM fields and in the presence of the radiation reaction, we may draw some conclusions on how the electron dynamics evolve with increasing EM field power. We plot \eqref{p_RR_long} and the solution of \eqref{p_RR_tr} in Figure~\ref{fig:classical}(b), revealing two intersecting curves. The value of $a_0$ at the intersection point scales as $a_0=a_\mathrm{th}\sim \epsilon_\mathrm{rad}^{-1/3}$, which corresponds to the threshold for the radiation effects to become dominant in the transverse motion of an electron in a EM field \eqref{long_motion}. So, for electron beam interaction with MCLP with $a_0<a_\mathrm{th}$, the emission of photons is mostly directed forward with electrons loosing their energy down to \eqref{p_RR_long}, but not scattering to the sides. For $a_0>a_\mathrm{th}$, the electrons will lose energy until the point where their transverse motion becomes important and re-acceleration by the field can occur. In strong field this motion is characterized by equation~\eqref{p_RR_tr}, which defines such electron motion, when the electron emits the same amount of energy as it absorbs from the field per cycle. Moreover, the trapping phenomenon, which is connected to the existence of stable and quasi-stable trajectories in the field, is only possible for $a_0>a_\mathrm{th}$. Thus, $a_\mathrm{th}\sim\epsilon_\mathrm{rad}^{-1/3}$ is the threshold between pure depletion and re-acceleration.

In what follows we test these conclusions in numerical simulations, employing nQED formalism, \emph{i.e.}, the probabilistic nature of the photon emission by an electron and electron-positron pair production by a photon in a strong EM field. It can be expected that, at relatively low peak intensities of the MCLP configuration, an electron will lose part of its initial energy upon interacting with the EM field, but will mainly go through the field and emit photons along the direction of its propagation. At increasingly higher intensities electrons will be scattered more strongly sideways, which will also be reflected in the photon emission directions. At very high intensities electrons will be scattered in all directions and, therefore, the photons will be emitted in a $4\pi$ angle. The energy of the electrons will in this case be characterized by equation~\eqref{p_RR_tr}. Furthermore, $a_\mathrm{th}$ should determine the optimal value for the energy depletion of the electron beam, since $a_\mathrm{th}$ is defined as the field strength at which the electron momentum acquires its minimum value.

\begin{figure*}[t!]
	\includegraphics[width=0.8\textwidth]{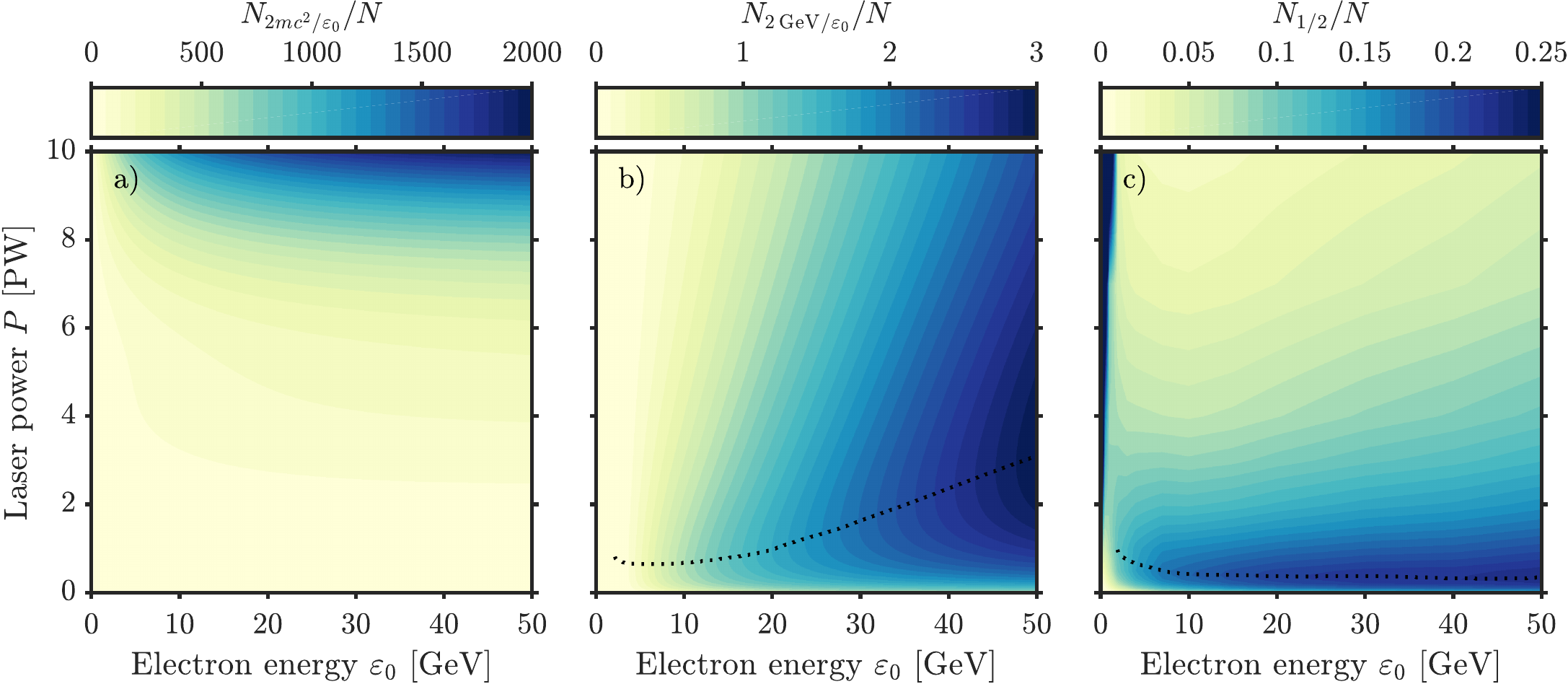}
	\caption{Total number of photons, normalized to the initial number of electrons in the beam ($N_{\varepsilon_\mathrm{th}/\varepsilon_0}/N$), detected above an energy threshold $\varepsilon_\mathrm{th}$ of (a) $2mc^2$, (b) $\unit[2]{GeV}$ and (c) $\varepsilon_0/2$ at the end of the simulation. (b)-(c) The dotted lines indicate the optimal laser power as a function of electron energy.}
	\label{fig:Nph}
\end{figure*}

\section{Interaction according to QED-PIC}\label{sec:QED-sims}
To verify the obtained predictions and to analyze the interaction beyond the made assumptions we turn to simulations based on the QED-PIC code \textsc{ELMIS} \cite{gonoskov.pre.2015}. We perform a set of simulations in which we vary the total incoming laser power $P$ and the initial energy $\varepsilon_0$ of electrons passing through the dipole wave. The dipole wave contains equal parts electric and magnetic modes with no phase difference, as in Section~\ref{sec:field}. It is generated at the boundary of the simulation box with a wavelength of $\lambda = \unit[1]{{\mu}m}$ and is supplied a constant (cycle-averaged) power. The dipole field is then formed through the self-consistent field evolution, taking into account potential suppression due to pair cascades. The simulation box was divided into $128 \times 128 \times 128$ cells of $(\lambda/16)^3$ resolution, giving the simulation box a physical size of $\unit[8]{{\mu}m}\times\unit[8]{{\mu}m}\times\unit[8]{{\mu}m}$.

Electrons were injected into the simulation box in the form of a Gaussian shaped electron bunch with a $L = \unit[5]{{\mu}m}$ FWHM longitudinal and $w = \unit[1]{{\mu}m}$ FWHM transverse size. The bunch is centred on and travels along the dipole axis with a total charge of $\unit[100]{pC}$, corresponding to $N = 6.2\times 10^8$ electrons. These values are similar to what is expected of electron bunches obtained from LWFA (see Ref.~\cite{gonsalves.prl.2019} and references therein).

As the electrons interact with the EM field photon emission and electron-positron pair creation is accounted for stochastically through the rates of nonlinear Compton and multi-photon Breit-Wheeler. To keep the simulations computationally feasible, photons with an energy below $2m_ec^2$ are discarded as their chance of pair creation is minimal and therefore do not further contribute to the interaction dynamics. A particle thinning routine \cite{gonoskov.arxiv.2016} is also employed in order to deal with the onset of pair cascades. The interaction is analyzed using data gathered on all particles inside the main interaction region, defined as a sphere of radius $\unit[3]{{\mu}m}$ centered at the dipole focus, and statistics is also captured on all particles leaving it. This data constitutes the main results of this study and is described in more detail in the following subsections.

The field structure described in Section \ref{sec:field} corresponds to a dipole pulse driven at a constant power. Since physical laser pulses are of finite energy and duration, the field structure can only be approximately described by equation (\ref{eq:EM_phi}-\ref{eq:EM_z}) for sufficiently long pulse durations. For more general temporal shapes, the field structure can most easily be obtained by generating the (incoming) field in the far-field region ($R \gg 1$) and then numerically advancing this field according to Maxwell's equations using a standard field solver (for analytical consideration of this problem see Ref.~\cite{gonoskov.pra.2012}). Thus, the electric mode of the dipole field can be generated in the far-field region using
\begin{align}
\vec{B}^{ED} &= \frac{3a_0}{2R} {\sin(t+R)\Psi(t+R)} [\hat{n}\times\hat{d}], \\
\vec{E}^{ED} &= \frac{3a_0}{2R} {\sin(t+R)\Psi(t+R)} \big[\hat{n}\times[\hat{n}\times\hat{d}]\big],
\end{align}
where $\Psi(t+R)$ describes the temporal envelope of the pulse, $\hat{n}=\vec{R}/R$ and $\hat{d}$ is the (normalized) dipole vector, and similarly for the magnetic mode.

\subsection{Source of high energy photons}\label{sec:photons}
For the purpose of discussing the capabilities of the proposed source we define a generation efficiency measure based on the total number of photons $N_x$ that escapes the interaction region with an energy above a given threshold ($\varepsilon_\mathrm{th} = x\varepsilon_0$). In regimes where re-acceleration is weak an electron can at most emit one photon with an energy above $\varepsilon_0/2$, making $N_{1/2}/N$ a natural measure for high-energy photon generation as it also becomes bounded by $100\%$.

The photon generation efficiency is presented in Figure \ref{fig:Nph} as a function of laser power and electron beam energy for three different threshold energies. For low thresholds, Figure \ref{fig:Nph}(a), the measure effectively gives the total number of photons and shows that it increases with both laser power and electron beam energy, as could be expected. For large thresholds, Figure \ref{fig:Nph}(b), this trend is broken and an optimal laser power, for the purpose of high-energy photon generation, emerges. The explanation for this is found by considering the interplay between the Compton and Breit-Wheeler processes that both increase in rate with increasing field strengths. While this interplay initially favours an increasing production of high-energy photons the fraction of high-energy photons able to escape the interaction region will eventually, for sufficiently large field strengths, starts to decrease and instead fuel a shower-type cascade \cite{mironov.pra.2014}. 

The efficiency for generating photons above half the initial electron beam energy ($\varepsilon_\mathrm{th} = \varepsilon_0/2$) is shown in Figure~\ref{fig:Nph}(c). Similarly to Figure~\ref{fig:Nph}(b), there is a region in which the efficiency initially increases with increasing laser power, but eventually drops off as the photon decay into pairs begins to dominate. What is even more interesting is that the optimal laser power is almost completely flat around a relatively moderate $\unit[0.4]{PW}$ for a large range of electron energies, with an efficiency of $\sim 20\%$. The efficiency was more generally shown in Ref.~\cite{magnusson.arxiv.2018} to depend on the field shape and field structures with the greatest intensity gradients where shown to be superior.

The flatness in optimal power in Figure~\ref{fig:Nph}(c) can also be explained by analyzing the probability of the emission of high-energy photons through the Compton process. The differential emission rate in the case of $\chi\ll 1$ and $\varepsilon_\gamma\rightarrow 1$ is \cite{bulanov.pra.2013}
\begin{equation}
dP^e=\frac{\alpha}{\pi\lambda_C}\frac{m_e}{\gamma}\frac{\chi_e^{1/2}}{(1-\varepsilon_\gamma)^{1/2}}\exp\left[-\frac{2}{3}\frac{\varepsilon_\gamma}{(1-\varepsilon_\gamma)\chi_e}\right]d\varepsilon_\gamma,
\end{equation}
where $\varepsilon_\gamma$ is the emitted photon energy normalized to its maximum value. The dependence of the number of high-energy photons on electron initial energy can then be calculated as $P^e_{1/2}(\gamma)=\int_{\varepsilon_\gamma/2}^{\varepsilon_\gamma} dP^e$, which demonstrates an almost flat behaviour for $\unit[0.4]{PW}$ laser power and electron energies in the $10$-$\unit[100]{GeV}$ range.

The second high efficiency region that can be seen in Figure~\ref{fig:Nph}(c), for $\varepsilon_0 \lesssim \unit[1]{GeV}$, is a direct result of reacceleration, which makes multiple emission of photons with an energy $>\varepsilon_0/2$ possible. While this regime lends itself to larger numbers of such photons their energies are, as previously discussed, relatively low as they are limited to only a few GeV.

\begin{figure}[t!]
	\includegraphics[scale=0.4]{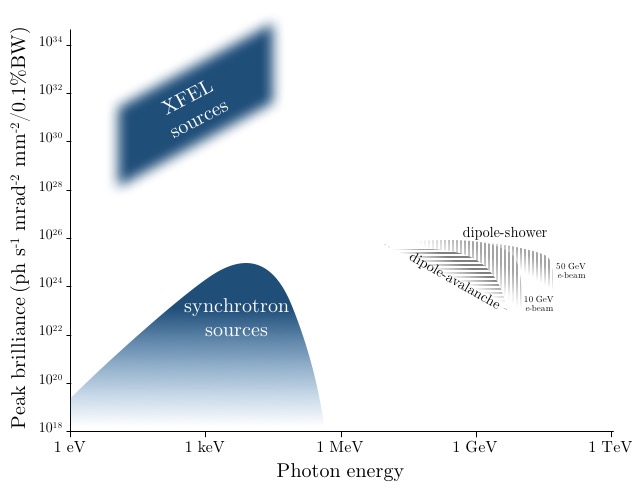}
	\includegraphics[scale=0.4]{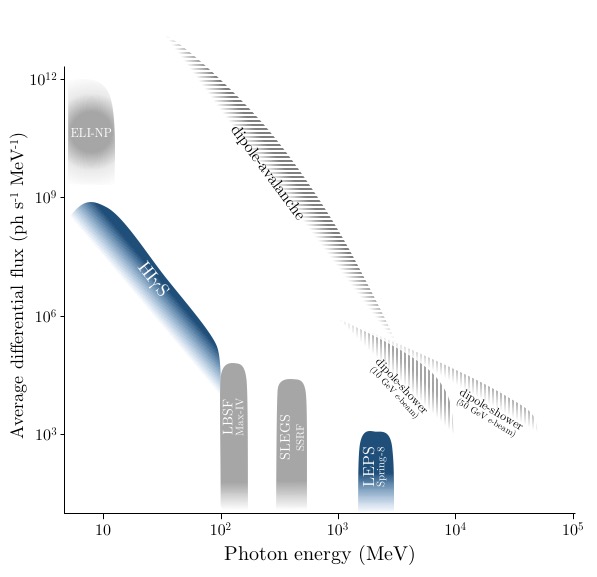}
	\caption{Comparison of the dipole-source with existing sources, in terms of (a) peak brilliance and (b) average flux. The plots schematically illustrate the properties of existing (blue) as well as upcoming or planned (solid gray) facilities. The gray hatched areas show the prospective properties of the dipole-avalanche source (horizontal) and our proposed source (vertical), the dipole-shower source.}
	\label{fig:sources}
\end{figure}

In Figure~\ref{fig:sources} we compare the capabilities of the source to both existing and previously proposed sources. Assuming an optimal laser power of $\unit[0.4]{PW}$ and a $\unit[10]{GeV}$ or $\unit[50]{GeV}$ electron beam, and with a physical size and total charge as specified in Section~\ref{sec:QED-sims}, the source is shown to be able to provide a peak brilliance, Figure~\ref{fig:sources}(a), similar to that of the avalanche-cascade source \cite{gonoskov.prx.2017} but at greater photon energies. For $\unit[10]{GeV}$ ($\unit[50]{GeV}$) the peak brilliance is estimated to $\unit[10^{26}]{ph/s\,mrad^{2}\,mm^{2}\,0.1\%BW}$ at $\unit[250]{MeV}$ ($\unit[370]{MeV}$). In comparing it to other Compton sources it is also instructive to compare them in terms of average flux Figure~\ref{fig:sources}(b). Assuming a repetition rate of $\unit[1]{Hz}$ the proposed source is comparable to current facilities in terms of average flux, but is able to reach greater energies. While these aspects allow us to compare the source to others it fails to capture what is perhaps the most interesting aspect of the dipole-sources, namely, that the high-energy photons are produced in dense bunches on the order of $\unit[10^{20}]{cm^{-3}}$.

\subsection{Onset and occurence of pair-cascades}\label{sec:pairs}
When an energetic electron beam interacts with a strong EM field its energy is converted into secondary electrons, positrons and photons in a process commonly referred to as a cascade, which can be of either shower- or avalanche-type. The shower-type corresponds to the case where the initial energy of particles (or photons) entering the strong field region is driving the cascade, whereas the avalanche-type corresponds to the case where the energy of particles participating in the cascade is continuously replenished through their repeated acceleration by the same strong field. Because both types involve a production of electron-positron pairs, characterizing the interaction by the number of pairs produced can therefore give an indirect estimate of the strength of the cascade.

For low initial energies, the electron dynamics is dominated by the influence of the much stronger field ($\gamma \ll a_0$). As the electrons rapidly lose their initial inertia they can get trapped in this field, where they produce pairs through their re-acceleration followed by emission of high-energy photons, initiating an avalanche-type cascade. Due to the probabilistic nature of the QED processes it becomes possible to observe a notable yield in high-energy photon production well below the values of $P$ required for the dominant appearance of the trapping. As this trapping becomes self-sustained the comparison becomes formally inconsistent because the number of particles produced inside and emitted out from the dipole-field continues to grow until the end of the simulation, necessitating a comparison that takes the finite duration of the laser field into account. In the case of an electric dipole, the avalanche-cascade becomes self-sustained for $P_\mathrm{min} \simeq \unit[7.2]{PW}$ \cite{gonoskov.prx.2017}, but for the electromagnetic dipole studied here this threshold is increased to beyond $\unit[10]{PW}$. This fact is evident in the absence of a significant pair production in conjunction with the efficient photon production. Nevertheless, as will made apparent in the subsequent discussion, this regime exhibits most other features of a cascade, leading us to signify it as an \emph{avalanche precursor}.

\begin{figure}[t!]
\includegraphics[width=0.5\textwidth]{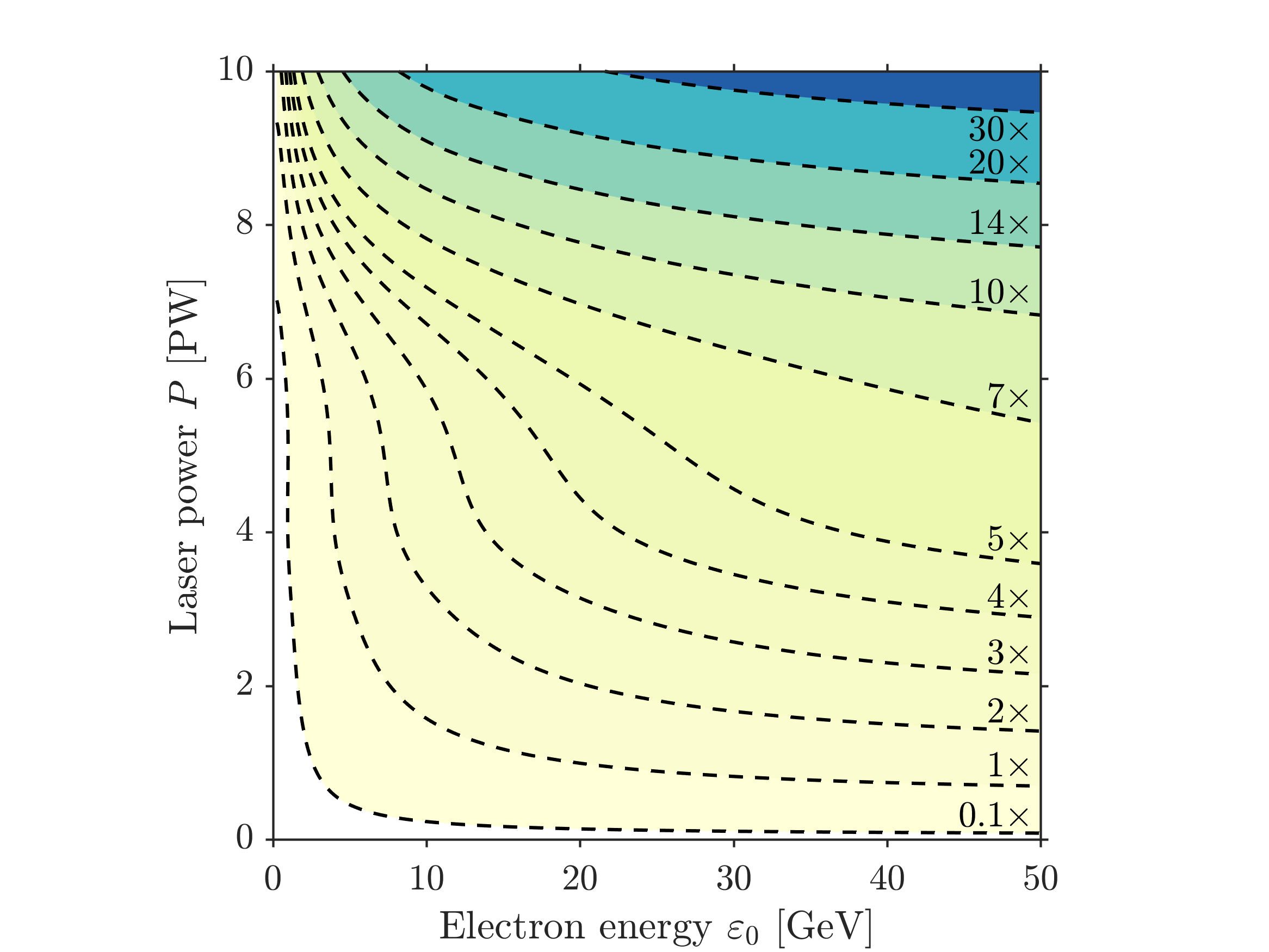}
\includegraphics[width=0.5\textwidth]{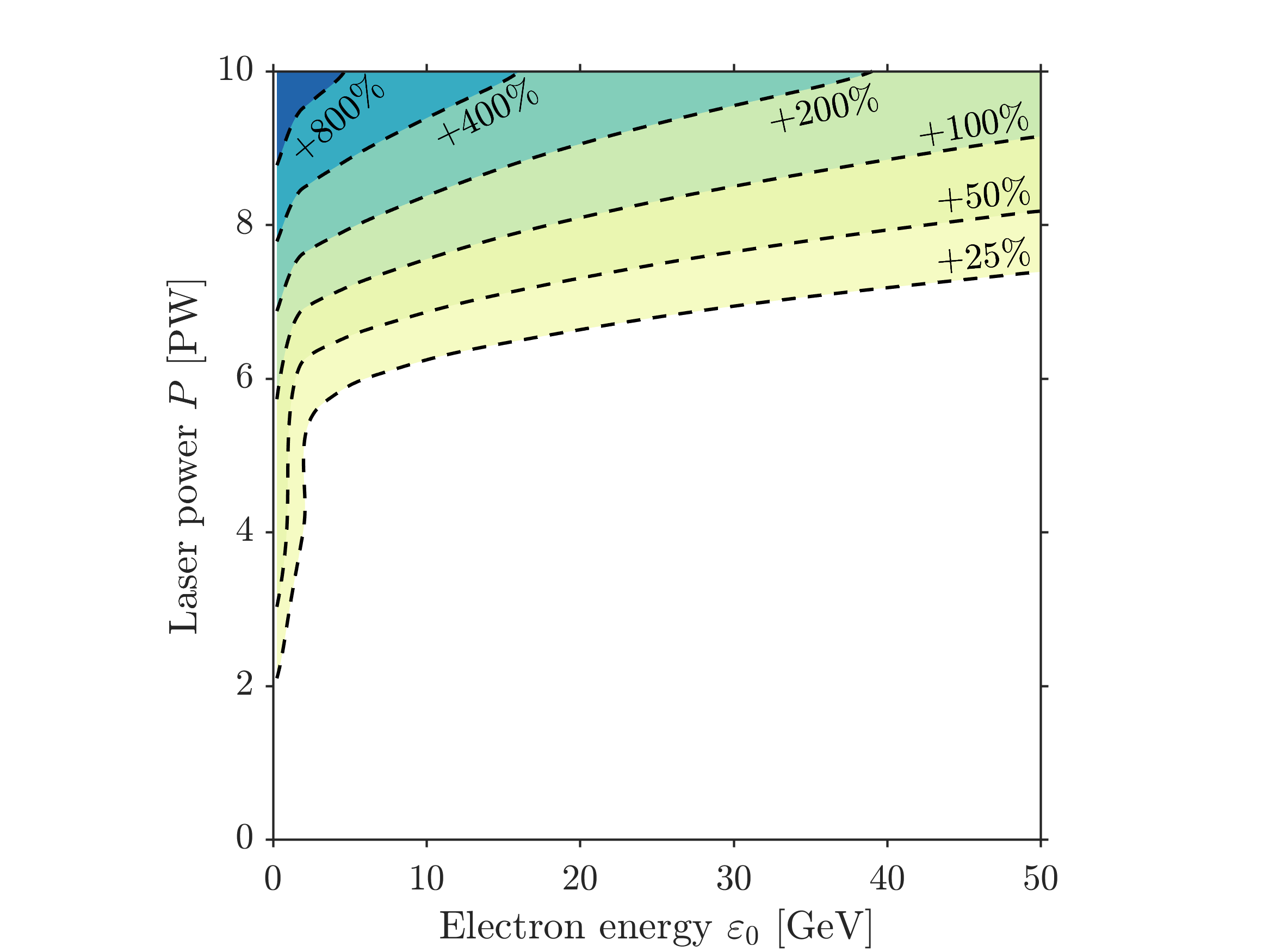}
\caption{The dependence of (a) the number of generated electron-positron pairs relative to the initial number of electrons in the beam ($N$) and (b) the total energy carried by all particles at the end of the simulation relative to the initial energy carried by the electrons in the beam, on the initial electron energy and laser power.}
\label{fig:Np}
\end{figure}

For larger values of initial energy ($\gamma \gtrsim a_0$) this avalanche precursor no longer appears. This is because both the injected and the produced particles particles have a large enough inertia to penetrate through the high-field region before depleting their energy and get trapped. As was discussed in Section~\ref{sec:photons}, the onset of a shower-cascade ultimately hampers the yield of high-energy photons in favour of an increasingly larger production of pairs. This becomes clear upon comparing Figure~\ref{fig:Nph}(b)-(c), showing the number of high-energy photons produced, with Figure~\ref{fig:Np}(a), showing the total number of produced pairs. While the high-energy photon production for $\varepsilon_0 \gtrsim \unit[1]{GeV}$ and $P \gtrsim \unit[1]{PW}$ drops with increasing laser power, an increase in either parameter consistently result in a larger number of electron-positron pairs. The regimes of efficient high-energy photon production and efficient pair production are, therefore, clearly separated in the parameter space and it becomes possible to generate as many as $20$ ($40$) times the initial number of electrons, at $\unit[10]{PW}$ and $\unit[10]{GeV}$ ($\unit[50]{GeV}$).

Neglecting the effects of transverse motion and assuming that the electron dynamics is fully determined by the depletion of the electron energy, then the sum of the energy of all particle species is approximately equal to the initial beam energy. This is in principle how the notion of the shower-type cascade is defined, that the energy supplied to the cascade comes almost exclusively from the initial particle beam. In our simulations we tracked the evolution of the total energy carried by each particle species. It was found that for most values of initial electron energy and laser power the sum of these energies at the end of the simulation is approximately equal to the total energy initially carried by the electron beam. However, when the laser power approaches $\unit[10]{PW}$ this changes significantly, see Figure~\ref{fig:Np}(b). While the combined energy of all the electrons and positrons at the end of the interaction is either approximately equal to or smaller than the energy initially carried by the beam, for most values of $\varepsilon_0$, the total energy carried away by the emitted photons can significantly exceed that of the initial beam. This becomes especially pronounced for lower values of $\varepsilon_0$, with $12$ times the initial energy being emitted as photons in the case of $\unit[1]{GeV}$ compared to $2.6$ times the initial energy in the case of $\unit[50]{GeV}$ (both at $\unit[10]{PW}$). This indicates that a significant re-acceleration begins to develop as we approach $\unit[10]{PW}$, especially at low $\varepsilon_0$, as the added energy can only come from the laser field.

\begin{figure*}[t!]
\includegraphics[width=1\textwidth]{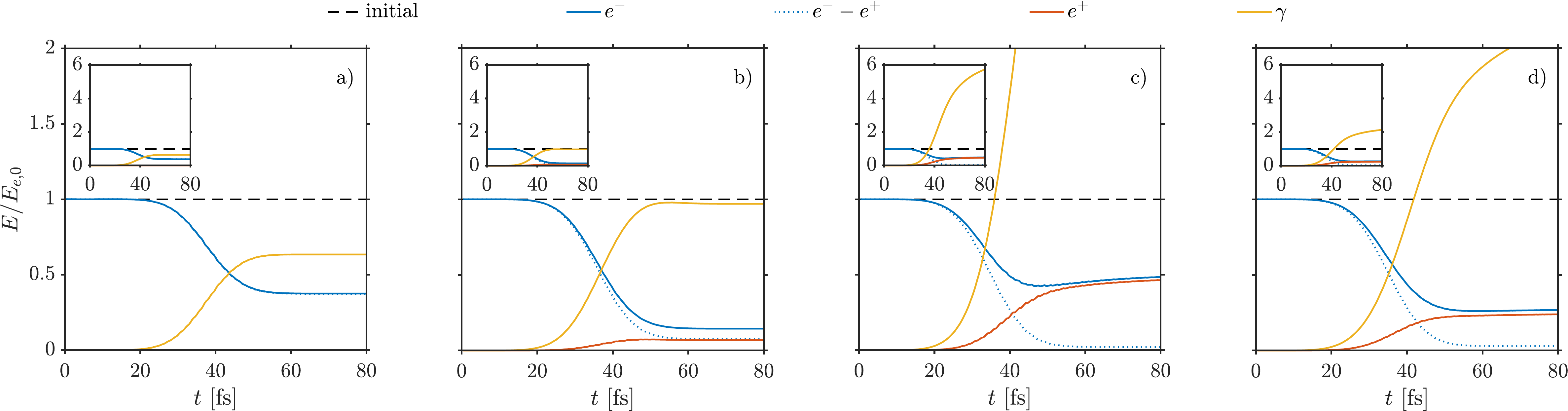}
\caption{The temporal evolution of the total energy, partitioned between electrons (blue), positrons (red), and photons (yellow) for different values of laser power $P$ and initial electron beam energy $\varepsilon_0$: (a) $\unit[1]{PW}$, $\unit[1]{GeV}$, (b) $\unit[4]{PW}$, $\unit[4]{GeV}$, (c) $\unit[10]{PW}$, $\unit[10]{GeV}$, (d) $\unit[10]{PW}$, $\unit[50]{GeV}$. The total energy of the initial electron beam (dashed) and the total energy carried by the electrons originating from the initial beam (dotted blue) is also shown. Insets represents the same data with a greater vertical scale.}
\label{fig:energypartition}
\end{figure*}

To further strengthen this view we may look at the energy partitioning between the different species at the end of the simulation, shown in Figure~\ref{fig:energypartition}. At $\unit[1]{GeV}$ and $\unit[1]{PW}$ (Fig.~\ref{fig:energypartition}(a)) the electron beam loses $61\%$ of its energy in the interaction such that the electrons and photons in the end carry $0.39 E_{0}$ and $0.62 E_{0}$, respectively, while the energy of the positrons is negligible. This means that the electron beam energy is the predominant source of energy, while a small additional amount comes from the acceleration of particles in the EM field.
A similar picture can be seen At $\unit[4]{GeV}$ and $\unit[4]{PW}$ (Fig.~\ref{fig:energypartition}(b)). Here the electron beam loses $94\%$ of its energy in the interaction, while photons and electron-positron pairs in the end carries $0.97E_{0}$ and $0.14 E_{0}$, respectively. The final energy, therefore, amounts to $1.2 E_{0}$, where the additional $0.2 E_{0}$ comes from the EM field. As we approach $\unit[10]{PW}$ the interaction becomes qualitatively different, with significantly more energy carried by photons than by electrons and positrons. Furthermore, while the total photon energy quickly saturates for lower values of $P$, this saturation gets significantly delayed around $\unit[10]{PW}$ and the emitted photon energy can be seen to increase until the very end of simulation at $\unit[80]{fs}$.

\begin{figure*}[t!]
	\includegraphics[width=1\textwidth]{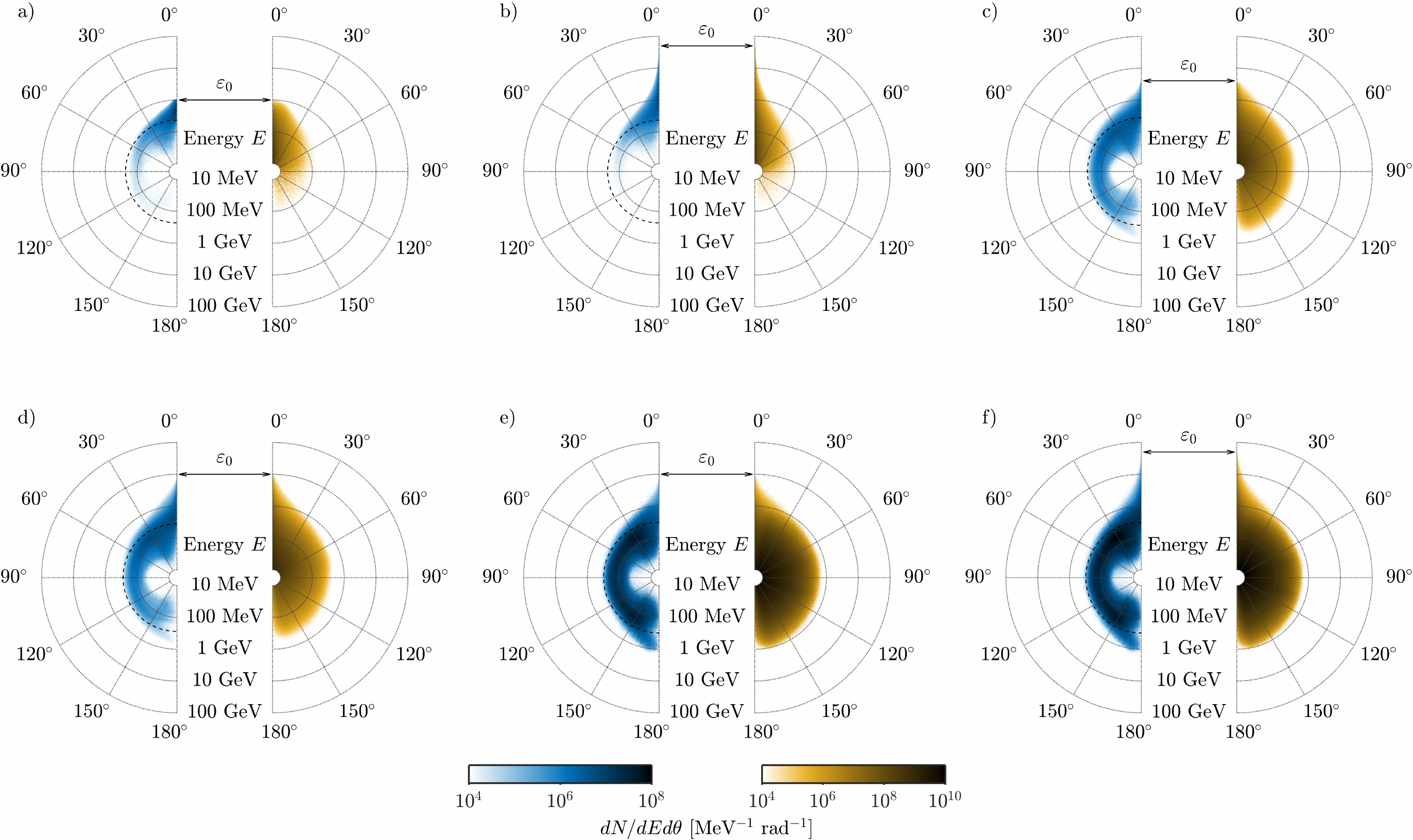}
	\caption{The energy-angle distributions of electrons (blue) and photons (yellow) emitted from the interaction for six cases of laser power $P$ and initial electron energy $\varepsilon_0$: (a) 1PW, 1GeV, (b) 1PW, 50GeV, (c) 4PW, 4GeV, (d) 4PW, 10GeV, (e) 10PW, 10GeV, (f) 10PW, 50GeV.}
	\label{fig:energyangle}
\end{figure*}

This change in behaviour is, as already mentioned, connected to a partial trapping of a few charged particles, which exceed the time it usually takes to escape strong field region. In this state, they move along quasi-stable orbits and radiate away the excess energy gained from the field, giving rise to the demonstrated extraction of energy from the field despite the absence of a fully self-sustained cascade. A qualitative explanation is provided by the analysis presented in Section~\ref{sec:classical_energy_loss}. If $a<a_\mathrm{th}$ then the electrons lose energy down to the value roughly given by equation~\eqref{p_RR_long}, with photon emission mainly directed forward. However, if $a>a_\mathrm{th}$ then the transverse motion of the electrons will become increasingly more important as they lose energy until its roughly determined by equation~\eqref{eq:energy_estimate}, where the losses are balanced by re-acceleration. Due to the transverse nature of this motion, photons (as well as pairs) will be emitted in all directions, as seen in Figure~\ref{fig:energyangle}, and with an energy roughly on the order of $m_e c^2(a/\epsilon_\mathrm{rad})^{1/4}$, as obtained in equation \eqref{eq:energy_estimate}.

\begin{figure*}[t!]
\includegraphics[width=0.75\textwidth]{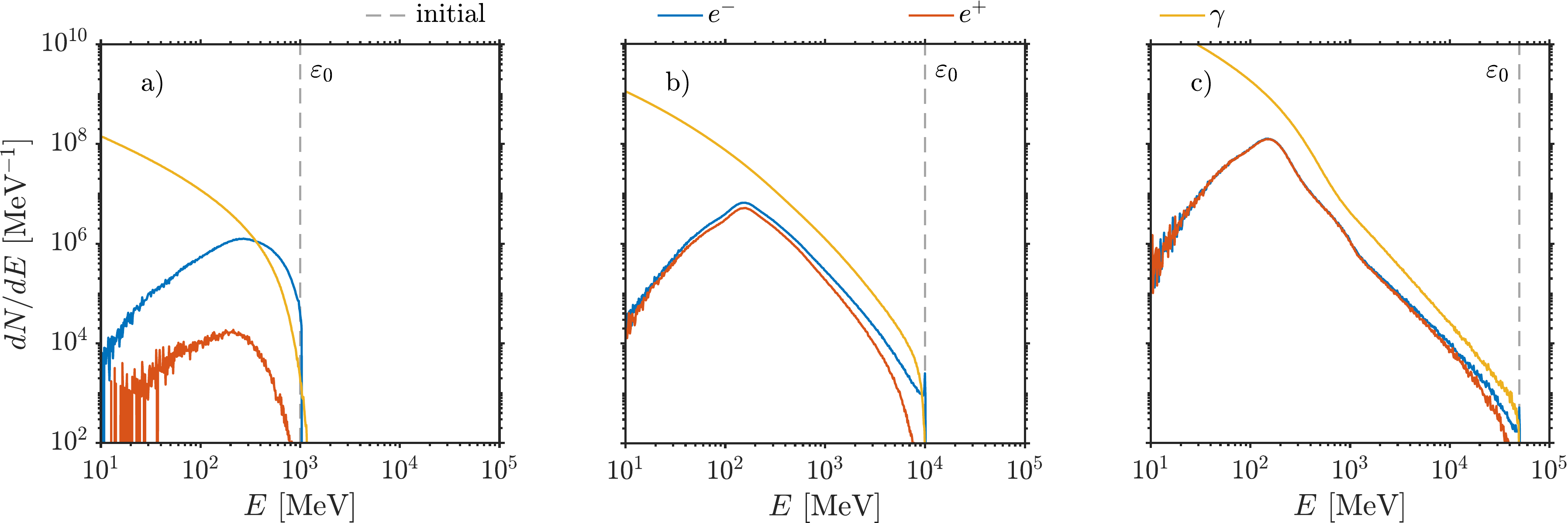}
\caption{The energy spectra of electrons (blue), positrons (red) and photons (yellow) at the at the end of the simulation for three different cases of laser power $P$ and initial electron energy $\varepsilon_0$: (a) 1 PW, 1 GeV,  (b) 4 PW, 10 GeV, (c) 10 PW, 50 GeV. Marked is also the initial electron energy $\varepsilon_0$ (dashed grey).}
\label{fig:spectra}
\end{figure*}

\subsection{Radiation dominated electromagnetic shield}\label{sec:radshield}
As shown previously in this section, a prolific production of photons occurs as the laser power approaches $\unit[10]{PW}$. These photons account for the largest contribution to the total energy of the final particles, which can significantly exceed the initial beam energy. At $\unit[10]{PW}$ the electrons and positrons, while temporarily trapped in the field, provide an effective mechanism for transforming laser energy into $\gamma$-ray photons. In Section~\ref{sec:classical_energy_loss} the energy loss of a high-energy electron injected into the field is studied on the basis of classical radiation reaction, suggesting that an electron beam can be effectively stopped over micron-size distances, as it radiates away most of its energy. This deceleration, according to equation \eqref{p_RR_long}, improves with increasing laser power. However, as we have seen, when the laser power is increased there is also a large production of photons and pairs. For this radiation shield to be considered effective it is therefore important that the generated particles have significantly lower energies than the electrons initially in the beam.

The expected electron energy as a function of longitudinal coordinate is presented in Figure~\ref{fig:classical}(a), showing a rapid energy loss over about a wavelength of propagation distance. The corresponding quantity from the simulations is the average energy of the initial beam electrons presented in Figure~\ref{fig:energypartition} (blue, dotted lines), but here as a function of time. Although these simulations were performed with an electron beam of finite length it shows a very similar trend, centred around $\unit[40]{fs}$, and where the broadening can explained by the $\sim\unit[15]{fs}$ duration of the electron beam. This general, monotonic trend can be seen to persist even in cases where the trapping of particles causes a significant generation of electron-positron pairs, as seen is Figure~\ref{fig:energypartition}(c).

\begin{figure}[t!]
\includegraphics[width=0.5\textwidth]{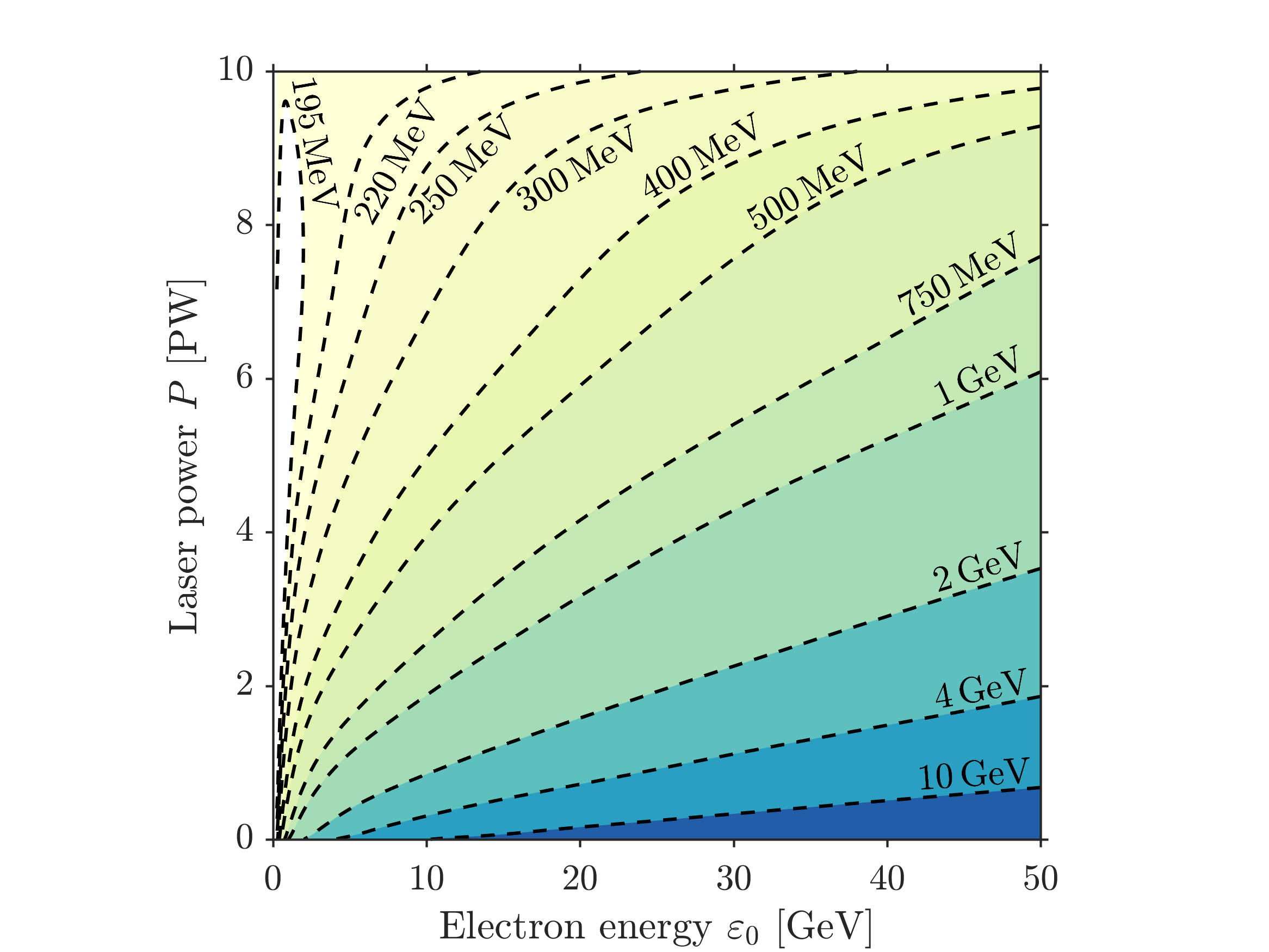}
\includegraphics[width=0.5\textwidth]{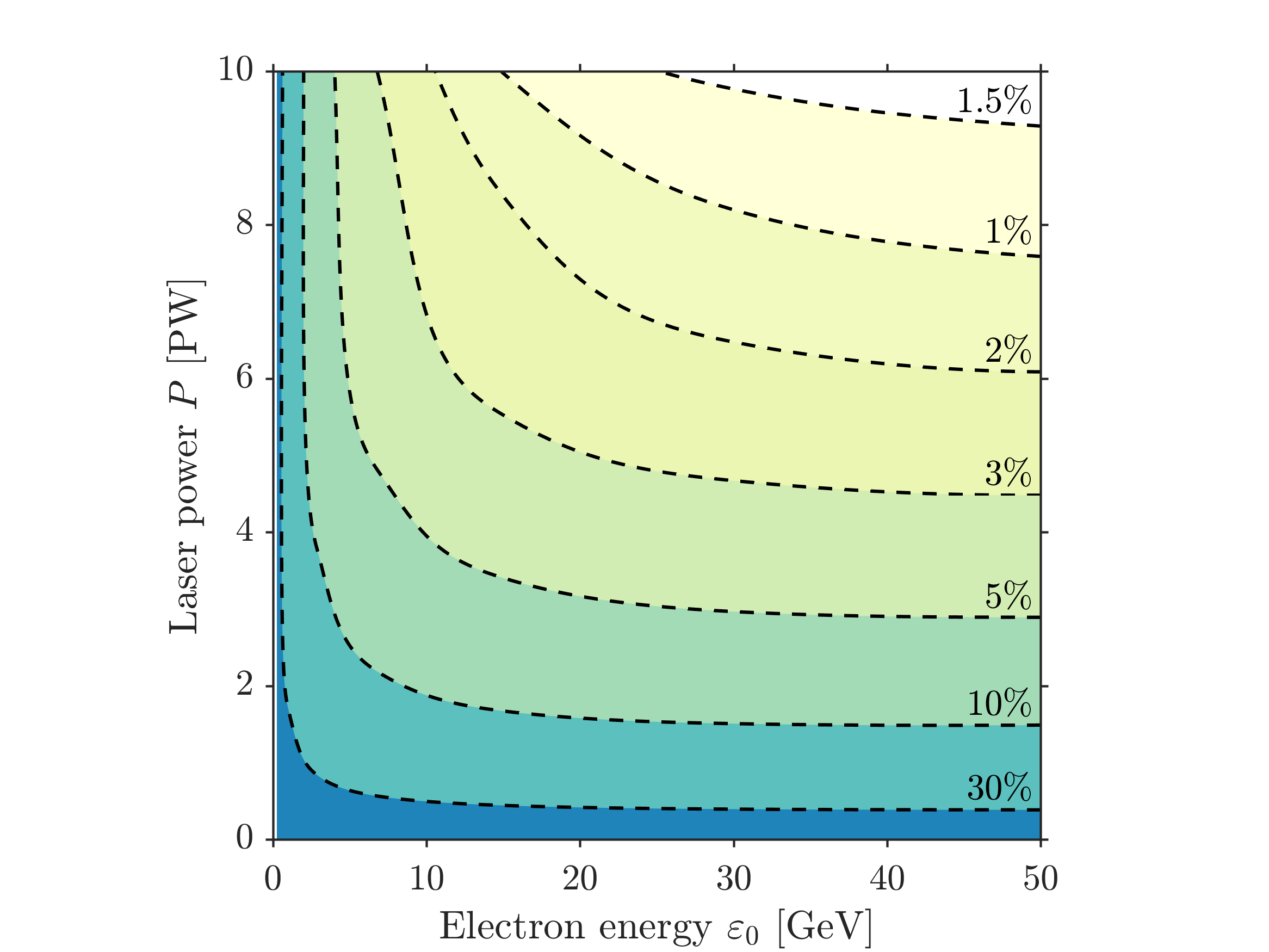}
\caption{The dependence of: (a) the average electron energy; and (b) the average electron energy relative to the initial electron energy $\varepsilon_0$;  on the initial electron energy $\varepsilon_0$ and laser power $P$ for all electrons.}
\label{fig:energyloss}
\end{figure}

The energy spectra of electrons, positrons and photons at the end of the simulation is shown in Figure~\ref{fig:spectra} for a few different laser powers and electron energies. While the total amount of energy carried by all species was earlier demonstrated to increase with laser power, the average energy of the electrons is here shown to at the same time decrease. This remains true for a large range of parameters and the electron and positron spectra can be seen to peak at lower values (below $\unit[200]{MeV}$) at high power, despite much greater initial electron energies. The final average energy (relative to the initial energy, $\varepsilon_0$) of electrons is also presented in Figure~\ref{fig:energyloss}(a) (Figure~\ref{fig:energyloss}(b)), further demonstrating how the average energy decreases with increasing laser power. The only deviation to this trend can be found at low $\varepsilon_0$, around the avalanche precursor, where the re-acceleration contributes to an increased average energy. 

\begin{figure}[t!]
\includegraphics[width=0.5\textwidth]{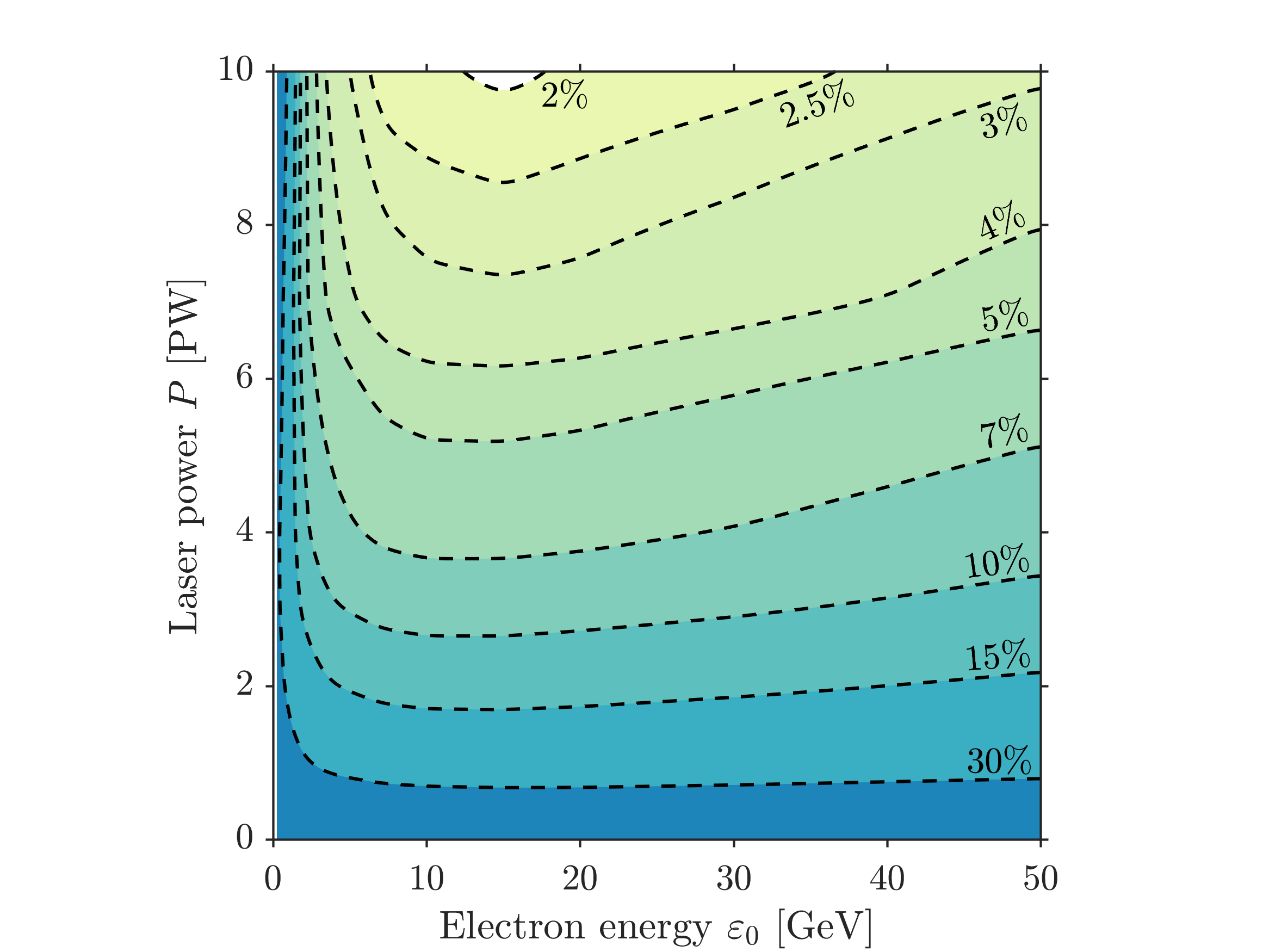}
\caption{The dependence of the average electron energy, relative to the initial electron energy, on the initial electron energy $\varepsilon_0$ and laser power $P$ for the electrons originating from the initial beam.}
\label{fig:energyloss2}
\end{figure}

Since the average electron energy for large laser powers becomes strongly dominated by the energy carried by the generated pair plasma, it is also instructive to look at the average electron energy of the electrons originating from the beam, excluding the electrons produced during the interaction, presented in Figure~\ref{fig:energyloss2}. While this figure in general shows the same trend as Figure~\ref{fig:energyloss}(b) there are mainly two things worth noting. First, the average energy of the beam electrons is greater than that of the entire ensemble, in practically all of the studied parameter space. Second, for a given laser power, the average energy can here be seen to increase with increasing initial energy, after a certain point. This is a direct consequence of the quantum suppression of radiation reaction. As the typical $\chi_e$ experienced by the electrons increases beyond unity, the average radiation loss scaling of $\chi_e^{2}$ is suppressed to $\chi_e^{2/3}$. The average radiation loss, as a ratio over the initial particle energy, therefore decreases with increasing $\varepsilon_0$, causing the shift in scaling seen in Figure~\ref{fig:energyloss2} at large $\varepsilon_0$.

In terms of numbers, a $\unit[1]{GeV}$ electron beam subject to a total MCLP laser power of $\unit[0.5]{PW}$ loses roughly $50\%$ of its initial energy, which is similar to the results of recent experiments on radiation reaction \cite{poder.prx.2018,cole.prx.2018}. This energy depletion rapidly increases with increased laser power, going as high as $98\%$ for $\unit[15]{GeV}$ and $\unit[10]{PW}$. While the energy losses decrease with increasing initial electron energy due to quantum suppression, this dependence is weak and the depletion is therefore almost constant for electron energies from $\unit[5]{GeV}$ to $\unit[50]{GeV}$, for fixed laser power.

\subsection{Electromagnetic field depletion}
At extremely high intensities the electron beam energy depletion discussed in Section~\ref{sec:radshield} may in principle be accompanied by depletion of the EM field, as detailed in Ref.~\cite{seipt.prl.2017}.  Over the course of the multiphoton Compton and Breit-Wheeler processes, which are responsible for transforming the initial electron beam energy into photons and electron-positron pairs, a significant number of photons are absorbed from the EM field. The interaction of an electron beam of sufficiently high charge with an intense EM field can therefore lead to depletion of the field energy. Given the high intensities achievable with the setup considered in this paper, it is not unthinkable for such a scenario to manifest itself. Based on the results of Ref.~\cite{seipt.prl.2017}, we estimate that for the field depletion to amount to a sizeable contribution to the interaction of a $\unit[10]{PW}$ MCLP field and a $\unit[500]{MeV}$ ($\unit[50]{GeV}$) electron beam, the number of particles in the beam should be $10^{13}$ ($10^{11}$). For a $\unit[100]{pC}$ electron beam in the energy range studied in this paper, the peak field of the MCLP setup must exceed the Schwinger field in order for the field depletion to become significant. Given the parameters considered here the interaction therefore not allow for field depletion to manifest itself.

\section{Discussion and Conclusions}

\begin{figure}[t!]
\includegraphics[width=0.5\textwidth]{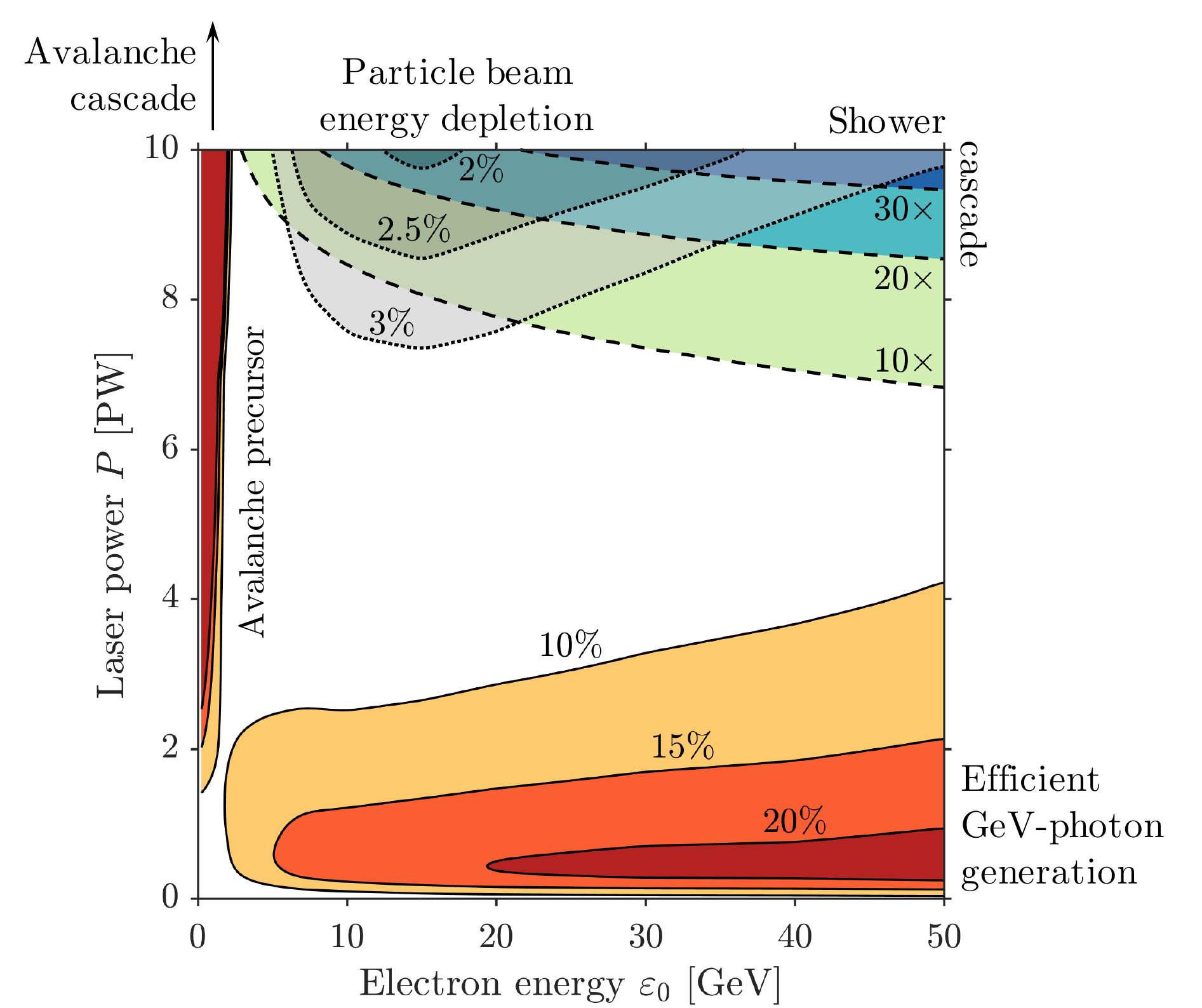}
\caption{The occurrence of different regimes in the interaction of a high energy electron beam with MCLP in the parameter space of initial electron energy $\varepsilon_0$ and total laser power $P$: (i) high-energy photon generation (contours of the number of high-energy photons normalized to the number of initial electrons, solid); (ii) shower-type cascade (contours of the number of e$^-$e$^+$ pairs normalized to the number of initial electrons, dashed); (iii) electron beam energy depletion (contours of the final electron beam energy as percentage of the initial electron beam energy, dotted). The contours of efficient high-energy photon generation occupy two separate regions: (1) the efficient generation of multi-GeV photons using electron beams of high energy; and (2) an avalanche precursor, characterized by a partial trapping and re-acceleration of particles, providing efficient generation of few-GeV photons.}
\label{fig:regimes}
\end{figure}

In this paper we studied the interaction of a relativistic electron beam with multiple-colliding laser pulses. This interaction is highly nonlinear and is further complicated by the dynamic interplay of strong field QED effects. We showed that depending on the interaction parameters, initial electron energy and laser power, the interaction can demonstrate a number of different regimes: (i) production of dense beams of high-energy photons, (ii) triggering shower-type cascades for reaching extreme states of generated electron-positron plasmas, and (iii) electron beam energy depletion. These regimes, summarized in Figure~\ref{fig:regimes} in a map over the studied parameter space, are the result of the balance in relative strength between the separate processes.

For low electron beam energies, the dynamics is shown to be dominated by the deflection and acceleration of the particles in the strong field. For increasingly greater field intensities, particles are trapped in the field focus for longer periods of time, giving rise to an avalanche precursor that becomes self-sustained at yet higher intensities, as discussed in Section~\ref{sec:photons}. The trapping provides an efficient source of both electron-positron pairs and high-energy photons but because the particle energies is limited by the re-acceleration provided by the field, this regime has a reduced efficiency for producing photons with energies above a few GeV. 

At high initial electron energies, multi-GeV photons can be efficiently produced. This efficiency is optimal around $\unit[0.4]{PW}$, as greater laser intensities shifts the balance between the Compton and Breit-Wheeler processes in favour of pair production, hampering the photon yield for the benefit of a shower-cascade. As shown in Ref.~\cite{magnusson.arxiv.2018}, the remarkable localization of the field provided by the dipole wave plays a crucial role in attaining this balance at low laser power and for achieving an overall high efficiency of the high-energy photon production. We here compared the capabilities of the source to existing and previously proposed sources, see Figure~\ref{fig:sources}. It is shown to be able to provide a peak brilliance and average flux similar to conventional synchrotron sources, but at greater photon energies. Another important aspect of this source is the fact that the photons are generated in bunches of high density ($\sim\unit[10^{18}]{cm^{-3}}$ using LWFA bunches), clean from heavy charged particles and neutrons, naturally accompanying bremsstrahlung sources. It also provides the possibility to create beams with at least a partial polarization of photons, by letting the electron beam pass the focused field off-center such that the radial acceleration of electrons leads to the presence of a predominant polarization direction. This can be important for experiments, e.g. on Delbr\"uck scattering \cite{koga.prl.2017}.

Two more regimes can be identified at high intensities. First, the shower-type cascade, discussed in Section~\ref{sec:pairs}, occurs when the energy of an electron beam is predominantly transformed into electron-positron pairs. As the high energy electrons of the initial beam enter the dipole field, they begin to lose energy through multiple emissions of photons. These photons then decay into electron-positron pairs, which themselves have enough energy to emit photons. At about $\unit[8]{PW}$ the number of electron-positron pairs is an order of magnitude larger than the initial number of electrons and this number increases rapidly with both increasing laser power and initial electron energy. Second, the large emission of photons at high intensities suggests a fast depletion of the electron beam energy. In Section~\ref{sec:classical_energy_loss} the field structure is predicted to be able to effectively stop the electron beam over about a wavelength of propagation distance. As discussed in Section~\ref{sec:radshield}, the electron energy losses are maximized for high intensities and around $\unit[15]{GeV}$, reaching as high as $98\%$ of the initial energy, as the quantum suppression of radiation emission reduces the relative energy losses at sufficiently large initial electron energies.

To conclude, in this paper we showed that the high-field MCLP configuration together with high-energy electron beam provides not only an advantageous framework for studying strong field QED phenomena, but also demonstrates the ability of this configuration to advance possible applications, in particular, as a source of high-energy photons. We showed that the MCLP configuration makes it possible to study strong field QED phenomena starting from single emission processes to multi-staged ones involving significant transformation of laser energy into emerging electrons, positrons, and photons. This is why the outlined prospects of using the MCLP configuration, together with previous theoretical findings \cite{gonoskov.prl.2014, gonoskov.prx.2017, efimenko.scirep.2018, efimenko.pre.2019, bulanov.prl.2010b, gonoskov.prl.2013, gelfer.pra.2015, vranic.ppcf.2016, gong.pre.2017}, provide a strong motivation for implementing more advanced and challenging focusing geometries, such as the dipole wave, in order to make better use of the produced laser radiation at large-scale laser facilities.

\begin{acknowledgments}
J.\,M. would like to thank T.\,G. Blackburn for helpful discussions. S.\,S.\,B., C.\,G.\,R.\,G., C.\,B.\,S., and E.\,E. acknowledge support from the Office of Science of the U.S. DOE under Contract No. DE-AC02-05CH11231. J.\,K.\,K. acknowledges support from JSPS KAKENHI Grant Number 16K05639. S.\,V.\,B. acknowledges support at the ELI-BL by the project High Field Initiative (CZ.02.1.01/0.0/0.0/15 003/0000449) from the European Regional Development Fund. The research is partly supported by the Ministry of Education and Science of the Russian Federation under contract No.14.W03.31.0032 (A.\,G.), by the Swedish Research Council grants No. 2013-4248, 2016-03329 (M.\,M.) and 2017-05148 (A.\,G.), and by the Knut and Alice Wallenberg Foundation (A.\,G., J.\,M., M.\,M.). The simulations were performed on resources provided by the Swedish National Infrastructure for Computing (SNIC) at HPC2N.
\end{acknowledgments}

\bibliography{RadShield_long}{}
\bibliographystyle{aipnum4-1}

\vfill

\newpage

\end{document}